\documentclass[aps,prb,showpacs,reprint,groupedaddress]{revtex4-1}
\usepackage{color}
\usepackage{setspace}
\usepackage{verbatim}
\usepackage{graphicx}
\usepackage{subfigure}
\usepackage{amsmath}
\usepackage{soul}
\usepackage{multirow}

\definecolor{orange}{rgb}{1.0,0.76,0.02}

\setstcolor{blue}

\begin{document}

\title{Kinetic Monte Carlo simulations of vacancy diffusion\\ in non-dilute Ni-X (X=Re,W,Ta) alloys}

\author{Maximilian Grabowski}
\email{maximilian.grabowski@rub.de}
\affiliation{Interdisciplinary Centre for Advanced Materials Simulation, Ruhr-Universit{\"a}t Bochum, 44780 Bochum, Germany}
\author{Jutta Rogal}
\affiliation{Interdisciplinary Centre for Advanced Materials Simulation, Ruhr-Universit{\"a}t Bochum, 44780 Bochum, Germany}
\author{Ralf Drautz}
\affiliation{Interdisciplinary Centre for Advanced Materials Simulation, Ruhr-Universit{\"a}t Bochum, 44780 Bochum, Germany}

\begin{abstract}
The mobility of vacancies in alloys may limit dislocation climb.  Using a combined density functional theory and kinetic Monte Carlo approach we investigate vacancy diffusion in Ni-Re, Ni-W, and Ni-Ta binary alloys up to 10~at.\% solute concentration.  
We introduce an interaction model that takes into account the chemical environment close to the diffusing atom to capture the effect of solute-host and solute-solute interactions on the diffusion barriers. 
In contrast to an ideal solid solution it is not only the diffusion barrier of the solute atom that influences the vacancy mobility, but primarily the change in the host diffusion barriers due to the presence of solute atoms.
This is evidenced by the fact that the observed vacancy slowdown as a function of solute concentration is larger in Ni-W than in Ni-Re, even though Re is a slower diffuser than W. 
To model diffusion in complex, non-dilute alloys an explicit treatment of interaction energies is thus unavoidable.
In the context of Ni-based superalloys two conclusions can be drawn from our kinetic Monte Carlo simulations: the observed slowdown in vacancy mobility is not sufficient to be the sole cause for the so-called Re-effect; and assuming a direct correlation between vacancy mobility, dislocation climb, and creep strength the experimentally observed similar effect of W and Re in enhancing creep strength can be confirmed. 
\end{abstract}
\maketitle

\section{Introduction}
\label{sec:intro}
Solid-state diffusion is a key aspect in the evolution of the microstructure that ultimately determines many materials properties.
In multicomponent alloys atomic transport is of importance not only during processing where it influences the chemical composition of precipitates as well as the precipitation kinetics,~\cite{CLE06} but also under service conditions where it controls solute segregation to defects such as dislocations,~\cite{SF83,KLC18} grain boundaries,~\cite{HS77,HL96} interfaces,~\cite{HS77,RW89,DLD08} and surfaces.~\cite{MM78}
Diffusion also affects the stability of the microstructure and plays a role in processes such as dislocation climb,~\cite{MCF08} which in turn can be related to creep strength and other materials properties.~\cite{WBN14}  In Ni-based superalloys, a class of special purpose high-temperature materials,~\cite{Mee81} it was shown that the addition of a few at.\%~Re significantly enhances the creep strength, which is known as the Re-effect.~\cite{Ree06}  One of the hypothesis to explain the Re-effect is that Re hinders the transport of vacancies through the $\gamma$-channel and thus retards dislocation climb.~\cite{WBN14}
On the atomistic level the atomic jump frequencies that determine diffusion can be obtained with high accuracy by electronic structure calculations.  The jump frequencies can then be combined with analytical models~\cite{LL55,Nas05,MWA08,GNB13,GTN14} or numerical kinetic Monte Carlo (KMC) simulations~\cite{SRD14,GM14} to determine macroscopic diffusion properties.  Analytical models like the 5-frequency model~\cite{LL55} or the self-consistent mean field method~\cite{Nas05} are limited in the type of interactions that can be considered and only applicable within the dilute limit.  For larger solute concentrations analytical models only exist for the case of ideal solid solutions.~\cite{Man71,HE86,MAA89}
While in principle KMC simulations are not limited in their complexity regarding the alloy composition, in practice it may become an unfeasible task to determine the atomic jump frequencies accurately. A general way to include complex interactions in KMC models is to extract these interactions from electronic structure calculations and to parametrize a model lattice Hamiltonian.~\cite{ZS15,VCA01,BV15}

In this work we combine density functional theory (DFT) calculations with KMC simulations to investigate the influence of solute atoms on the mobility of vacancies in binary Ni-X (X=Re, W, Ta) alloys. We take into account the change in the atomic diffusion barriers due to the presence of multiple solute atoms explicitly.  This allows us to study compositions in the non-dilute limit.  The Ni-Re system serves as a model system for the $\gamma$-phase (face-centered cubic solid solution) in Ni-based superalloys.  A typical nominal Re concentration in Ni-based superalloys is around 1~at.\% which can be considered to be within the dilute limit. Both DFT based KMC simulations~\cite{SRD14} and analytical models~\cite{GM14} have shown that within the dilute limit the mobility of vacancies is only marginally reduced by about 4\% due to the presence of Re. In contrast, if one assumes that the observed increase in creep strength is due to retarded dislocation climb a vacancy slowdown of up to 75\% would be expected.~\cite{GM14}  
Due to segregation in the $\gamma/\gamma'$-microstructure of Ni-based superalloys the local Re concentration in the $\gamma$-channel can be significantly larger than the nominal one.
Energy-dispersive X-ray spectroscopy (EDX) and atom probe tomography (APT) experiments~\cite{PWB14} have shown that Re segregates to the $\gamma$-matrix and is almost non-existent in the $\gamma$'-particles. The local Re concentration can reach $8-10$~at.\% in the $\gamma$-phase, which is the non-dilute concentration range that we focus on in the current study.

In addition to Ni-Re we perform simulations in Ni-W and Ni-Ta.  Both W and Ta are further important alloying elements in Ni-based superalloys.  It was shown that a similar increase in creep strength as observed by the addition of Re can be achieved by increasing the concentration of W in the $\gamma$-phase.~\cite{FMA15}  Ta, on the other hand, partitions to the $\gamma'$-phase and its diffusion barrier in Ni is actually lower than the Ni self-diffusion barrier,~\cite{SRD14,GM14} i.e. Ta is a fast diffuser, which makes it interesting to compare to.

The interaction model that we introduce includes the most important contributions to the change in the diffusion barriers by evaluating the chemical composition in the vicinity of the diffusing atom.  
Even though our interaction model is limited in its complexity we observe a strong influence on the vacancy mobility as compared to an ideal random alloy.  This emphasizes the necessity to include the details of the atomic interactions when modeling diffusion properties in non-dilute alloys.

We briefly introduce our computational approach for the KMC simulations and set-up for the DFT calculations in Sec.~\ref{sec:compdetails}.  Next we present our DFT results for the diffusion barriers and introduce our interaction model (Sec.~\ref{sec:barriers}).  The vacancy mobilities extracted from the KMC simulations and a comparison of our interaction model with a random alloy are discussed in Sec.~\ref{sec:mobility}, before we conclude our results (Sec.~\ref{sec:conclusion}).

\section{Computational Approach}
\label{sec:compdetails}
\subsection{Kinetic Monte Carlo}
\label{subsec:kmc}
The mobility of the vacancies is evaluated using kinetic Monte Carlo simulations.~\cite{Gil76,BKL74,FW91}  Our KMC model represents vacancy-mediated substitutional diffusion on a face-centered cubic (fcc) lattice.  The rate constants $k_i$ for the exchange between a vacancy and a neighboring atom are given by harmonic transition state theory (hTST)~\cite{PW31,Eyr35,EP35} as
\begin{equation}
\label{eq:htstrate}
        k_i = \nu_i \exp\left(-\frac{\Delta E_i}{k_\text{B}T}\right)\quad,
\end{equation}
where $\Delta E_i$ represents the atomic diffusion barrier, $k_\text{B}$ is the Boltzmann constant, $T$ the temperature, and $\nu_i$ the attempt frequency.  Within hTST the attempt frequency is obtained from the vibrational frequencies in the initial and transition state of the diffusion process.  The diffusion barriers $\Delta E_i$ can be calculated using density functional theory (Sec.~\ref{subsec:DFT}).  A particular challenge in alloys with high solute concentrations is the dependence of the diffusion barriers on the local chemical environment.  A direct evaluation of the diffusion barriers using DFT during the KMC simulations is computationally unfeasible.  Likewise, an a priori list of all configurations cannot be established due to the vast amount of possible combinations.  One approach to capture the compositional dependence of the diffusion barriers within KMC simulations is to use a cluster expansion (CE)~\cite{SDG84} of the energy together with kinetically resolved activation (KRA) energies.~\cite{VCA01}  To obtain a fully converged CE of the energies as well as the diffusion barriers is a sophisticated task.  In the current work we focus on the key effects of multiple solute atoms in the vicinity of the diffusing atom and incorporate this dependence in our KMC simulations via a straightforward interaction model (Sec.~\ref{subsec:model}).

The vacancy tracer diffusion coefficient $D$ is extracted from the KMC trajectories using the mean square displacement $\left<R^2(t)\right>=N^{-1}\sum_i^N(r_i(t)-r_i(0))^2$, where $N$ is the number of vacancies (in this work $N=1$) and $r_i(t)$ is the position of vacancy $i$ at time $t$
\begin{equation}
 \label{eq:diffusion}
D = \frac{\left<R^2(t)\right>}{2dt} \quad ,
\end{equation}
where $d$ is the dimensionality.  The diffusion constant is measured over segments with a fixed number of KMC steps and the value of $D$ is determined from the time-weighted average over these segments.~\cite{DRD12}

The diffusion coefficient is frequently expressed as
\begin{equation}
        \label{eq:arrhenius}
        D=D_0\exp\left(-\frac{Q}{k_{\rm B}T}\right)\quad,
\end{equation}
where $D_0$ is the diffusion prefactor and $Q$ is the diffusion activation energy.  For vacancy-mediated substitutional diffusion $Q$ comprises the diffusion barrier and the vacancy formation energy for host and solute diffusion, whereas for vacancy diffusion $Q$ corresponds to the effective diffusion barrier.  The diffusion prefactor $D_0 = \Gamma a^2 \nu_0 f$ consists of a geometric factor $\Gamma = n/2d$ where $n$ is the number of possible jump sites and $d$ is the dimensionality, the jump distance $a$, the attempt frequency $\nu_0$, and the correlation factor $f$. For an fcc lattice $\Gamma = 2$, $f_{\rm host} = 0.7815$,~\cite{MT79} $f_{\rm vac}=1$, and $a = a_0/\sqrt{2}$ where $a_0$ is the equilibrium lattice constant. The analytic expression in Eq.~\eqref{eq:arrhenius} can be fitted to experimental as well as our KMC simulation data as a function of temperature to extract $Q$ and $D_0$.

The KMC simulations were performed using an extension to the CE code CASM.~\cite{CASM17}  The simulation cells contained 8788 sites including a single vacancy on an fcc lattice with periodic boundary conditions.  The solute concentration was varied between $2-10$~at.\% X (with X = Re, W, Ta) where the solute atoms were distributed randomly in the simulation cell to represent a solid solution.  The jump distance $a=2.48$~\AA~was derived from the equilibrium lattice constant of pure fcc Ni calculated using DFT.  The attempt frequency in Eq.~\eqref{eq:htstrate} was set to $\nu_0 = 10^{13}$~s$^{-1}$ which is a reasonable approximation in these systems.~\cite{GM14} All input values to our KMC model, in particular the composition dependent diffusion barriers $\Delta E_i$ in Eq.~\eqref{eq:htstrate}, were obtained from DFT calculations.
\subsection{Density functional theory calculations}
\label{subsec:DFT}
The DFT calculations were performed using the projector augmented wave (PAW) method~\cite{Blo94,KJ99} as implemented in the Vienna Ab initio Simulation Package (VASP 5.4),~\cite{KH93,KF96,KF96_2,KJ99}  with the gradient corrected PBE~\cite{PBE96} exchange-correlation functional.  All calculations were carried out spin-polarized as it was shown that magnetism significantly influences the interactions between solute atoms in Ni, in particular for Re.~\cite{HPG16}  
The diffusion barriers were determined using the climbing-image nudged-elastic band (CI-NEB) method~\cite{HUJ00,HJ00,JMJ98} as provided by the VTST package~\cite{VTST} for VASP.
Calculations were performed in $(3\times3\times3)$ fcc supercells with a $[4\times4\times4]$ Monkhorst-Pack~\cite{MP76} k-point mesh and a plane wave cutoff of 560~eV for the Ni-Re system and 500~eV for the Ni-W and Ni-Ta system.  Within this setup total energies were converged to within 1~meV/atom.  The ionic positions of the initial and final states of the diffusion process including a single vacancy were fully relaxed until forces were below $0.01$~eV~\AA$^{-1}$. The cell shape and volume were kept fixed at the equilibrium volume of $1158.88~$\AA$^{3}$, $1162.87~$\AA$^{3}$,  $1167.82~$\AA$^{3}$, $1170.87~$\AA$^{3}$,  $1174.87~$\AA$^{3}$, and $1178.88~$\AA$^{3}$ for one vacancy with $0-5$ Re atoms, respectively.
One image along the diffusion path was found to be sufficient for the CI-NEB calculations.  Forces were converged to $0.01$~eV \AA$^{-1}$ and the corresponding diffusion barriers to $10$~meV.

\section{Composition dependent diffusion barriers}
\label{sec:barriers}
\subsection{Re distribution in the diffusion channel}
\label{subsec:dft_nire}
\begin{figure}
        \includegraphics[width=5.0cm]{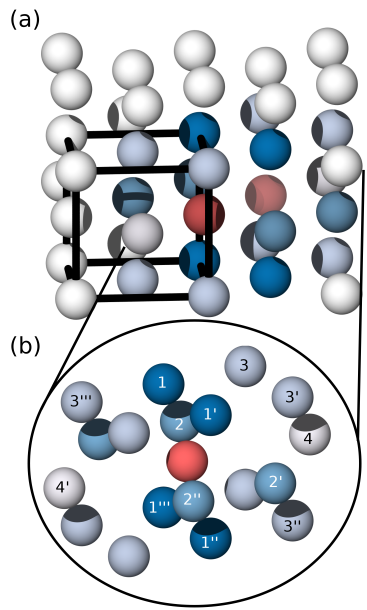}
        \caption{\label{fig:diffusion_channel} 
                        Fcc lattice depicting the diffusion channel (shades of blue) around a first nearest neighbor diffusion process. The diffusion channel consists of the first nearest neighbors around the initial and final position of the diffusing atom (red).
(b) The diffusion channel with 18 sites (blue) and the diffusing atom at the transition state (red). The notation indicates the distance from the transition state.
        }
\end{figure}
\begin{figure*}
        \includegraphics[width=\textwidth]{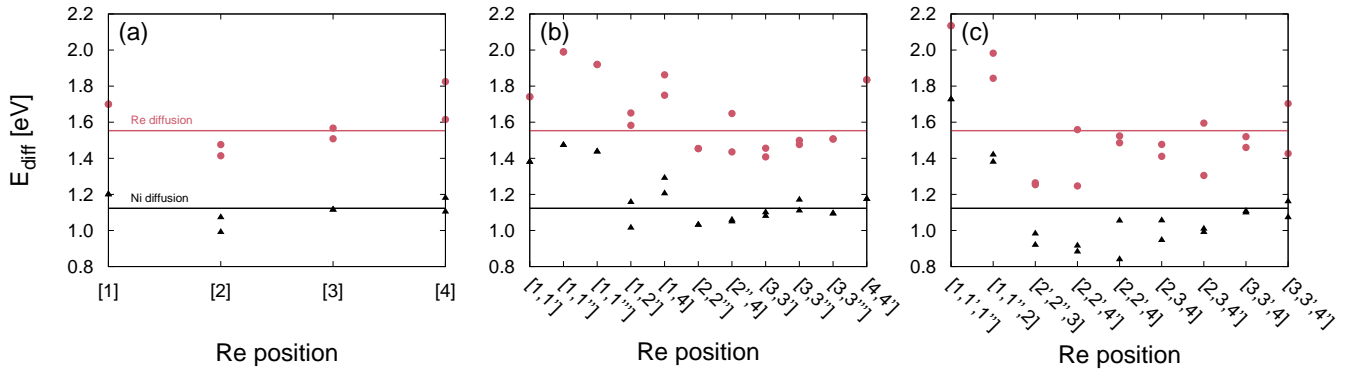}
        \caption{\label{fig:NN} Diffusion barriers with 1, 2, and 3 solute atoms in the diffusion channel at varying distances to the transition state. The primes denote the exact position in the diffusion channel as shown in Fig. \ref{fig:diffusion_channel}.  Configurations with $E_\text{IS}=E_\text{FS}$ have one value for the diffusion barriers and configurations with $E_\text{IS}\neq E_\text{FS}$ have two values, respectively.
}
\end{figure*}
The diffusion barriers in a binary alloy generally depend on the overall composition as well as the specific configuration of the initial (IS) and final (FS) state of the atomic diffusion process.  Usually changes in the local composition in the vicinity of the diffusing atom have the most significant effect on the barriers.
We therefore focus on different occupations of the diffusion channel shown in Fig.~\ref{fig:diffusion_channel} that comprise all first nearest neighbors of the initial and final position of the diffusing atom.  Here, we only consider first nearest neighbor jumps as diffusion processes.  Each of the 18 sites in the diffusion channel can be either occupied by Ni or Re, as well as the diffusing atom can be either Ni or Re.  Occupying 19 sites with two different elements yields $2^{19} = 524\,288$ possible combinations, and even if considering only symmetry nonequivalent configurations there are still too many possible structures to be directly calculated with DFT.  
As a model system for Ni-based superalloys we are only interested in local concentrations of Re of up to $\sim 10$~at.\%.  We therefore restrict the number of Re atoms in the diffusion channel to three, yielding a local concentration of $16.7$ and $22.2$~at.\% depending on if the diffusing atom is Ni or Re, respectively.
With three Re atoms in the diffusion channel there is still a large number of possible configurations from which we chose a set of structures 
where the distribution of Ni and Re in the diffusion channel is as dissimilar as possible so that these structures
represent a diverse variety of Ni-Re arrangements.  The set consists of 26 different occupations of the positions in the diffusion channel with $0-3$ Re atoms and one with 4 Re atom for which either Ni or Re can diffuse, yielding in total 52 structures.

The corresponding diffusion barriers are shown in Fig.~\ref{fig:NN}.  The $x$-axis denotes the different configurations where the index refers to the position of the Re atom in the diffusion channel relative to the transition state: $[1]$ means that the Re atom is located on a 1st nearest neighbor (NN) position to the transition state (see Fig.~\ref{fig:diffusion_channel}), $[2]$ that one Re is on a 2nd NN position, $[2'',4]$ that 2 Re atoms are located on the 2nd and 4th NN position, respectively.\\

Depending on the configuration the initial and final state of the diffusion process can either be equivalent or have different energies ($E_\text{IS} \neq E_\text{FS}$) due to interactions between the solute atoms and between the solute and the vacancy.  Correspondingly, the diffusion barrier will depend on the direction of the diffusion process with $\Delta E^\text{diff}_{\text{IS}\rightarrow \text{FS}}-\Delta E^\text{diff}_{\text{FS} \rightarrow \text{IS}}=E_\text{FS}-E_\text{IS}$.
The diffusion barriers for each configuration are shown in Fig.~\ref{fig:NN}; if $E_\text{IS}=E_\text{FS}$ there is only one value and if $E_\text{IS}\neq E_\text{FS}$ two values are given, respectively.
The straight lines indicate the diffusion barriers of Ni (black) and Re (red) in pure Ni with $\Delta E^\text{diff}_\text{Ni}=1.13$~eV and $\Delta E^\text{diff}_\text{Re}=1.55$~eV.  These values are in good agreement with previous calculations.~\cite{SRD14,GM14} 

Fig.~\ref{fig:NN}(a) shows the diffusion barriers with a single Re atom at the four nonequivalent NN positions to the transition state (TS) in the diffusion channel.  Re in the 1st NN position to the TS increases both the Ni and Re diffusion barrier by 80 and 140~meV, respectively, whereas  Re in the 2nd NN position slightly decreases the diffusion barriers and in the 3rd NN position has almost no effect.  Re in the 4th NN position has a larger impact on the diffusion barrier of Re than of Ni, which is mainly due to slightly attractive Re-Re interactions: in the IS the Re atom in the diffusion channel and the diffusing Re atom are 1st NN whereas in the FS the two Re atoms are separated, so that the energy of the IS is lower than the energy of the FS, while the energy of the TS is slightly increased by Re in 4th NN position.

A similar trend in the diffusion barriers is observed with two Re atoms in the diffusion channel (Fig.~\ref{fig:NN}(a)).  In particular with two Re atom in 1st NN position to the TS, $[1,1']$ and $[1,1'']$, the diffusion barrier significantly increases by $0.18-0.43$~eV for both Ni and Re.
Other configurations with at least one Re atom in 1st NN position also exhibit a general increase in the diffusion barriers, whereas Re in the 2nd NN position appears to reduce the barriers, due to a decrease of the TS energy.  
Re in the 3rd NN position to the TS only has a minor effect on the diffusion barriers.  Placing Re in 4th NN position seems to increase the diffusion barrier of Re, but not of Ni. This is again due to the slightly attractive 1st NN Re-Re interaction and a corresponding decrease in the IS/FS energies, while the TS energy slightly increases.

The diffusion barriers with three Re atoms in the diffusion channel shown in Fig.~\ref{fig:NN}(c) corroborate the general trend observed for one and two Re atoms:  Re atoms in 1st NN position to the TS strongly increase the diffusion barriers, Re atoms in 2nd NN position slightly decrease the diffusion barriers, and Re atoms in 3rd NN position have only a small effect.  In particular the configuration with three Re atoms in 1st NN position exhibits diffusion barriers that are $\sim 0.6$~eV  higher than in pure Ni for both Ni and Re.

Variations in the diffusion barriers arise from changes either in the IS/FS energies or in the TS energies.   
The difference in energy between the IS and FS is due to  Ni-Vac, Ni-Re, Re-Re, as well as Re-Vac interactions. Focussing to a first approximation on the Re-Re interactions we observe that configurations where three to four Re atoms are  second NN to each other usually lead to an energy gain of up to $400$~meV compared to the configuration where three to four Re atoms are placed as a first NN to each other.
Placing Re atoms on the third NN shell to each other has almost no effect on the energy of the IS and FS. However aligning  two Re atoms as first NN such that one Re atom can push another closer to the vacancy also exhibits a significant energy gain of up to $300$~meV. Configurations where Re atoms are  forth NN to each other usually correspond to an energy loss and are not very favourable. From this discussion it is already clear that Re-Re interactions in this ternary system are very complex and can only be addressed with a sophisticated cluster expansion approach.  In the current study we therefore focus only on the dominant contributions to a change in the diffusion barriers due the presence of multiple solute atoms.

Our DFT data suggests that the effect of Re in the diffusion channel on the TS energy is strongly dependent on the distance to the TS.  Most notably, Re in 1st NN position can be regarded as having a first order effect on the TS energy, whereas Re atoms further away from the TS only have second order effects.  This dependence has been discussed also for diffusion in the ordered L1$_2$ Ni$_3$Al phase.~\cite{LVP10} 
From Fig.~\ref{fig:diffusion_channel} it is intuitively clear why Re in 1st NN position to the TS has the largest influence on the TS energy.  The four 1st NNs form a window through which the diffusing atom has to move which leads to an increase in the TS energy when occupied with Re.  To a first approximation the influence of Re on the diffusion barriers can thus be described by considering the occupation of this \emph{diffusion window}.

\subsection{Diffusion window}
\label{subsec:model}
\begin{figure}
        \includegraphics[width=6.0cm]{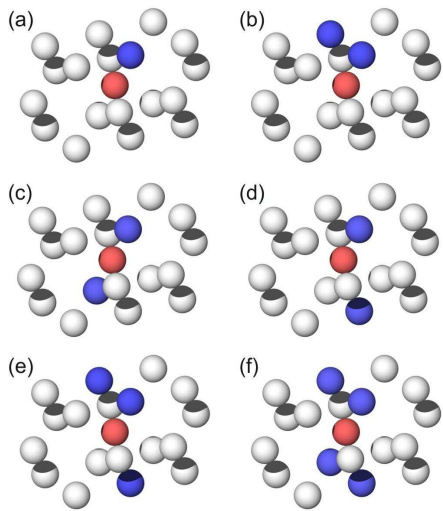}
        \caption{\label{fig:diffusion_channels} Simplified diffusion channels used in the KMC simulations. The diffusing atom is shown in red and the solute atoms in blue. Solute atoms can only occupy the 1st NN to the TS, the diffusion window. The simplest diffusion channel without any solute atom is not shown.
}
\end{figure}

Within the KMC simulations we locally determine the diffusion barriers that enter Eq.~\eqref{eq:htstrate} depending on the Re occupation of the 1st NN to the TS, i.e. of the diffusion window.  As discussed in Sec.~\ref{subsec:dft_nire} this allows us to capture the most important effect of Re on the diffusion barriers.  Re atoms in the diffusion window generally increase the diffusion barriers, whereas configurations that lead to a  decrease in diffusion barriers are not considered here.  Our model thus serves as an upper bound for the diffusion barriers in the presence of Re.
Except for the influence of Re in the diffusion window on the corresponding diffusion barriers we do not explicitly include any Re-Re interactions.  
Since Re-Re interaction energies are small~\cite{HPG16} and since we do not expect any clustering, this approximation is reasonable within the concentration range of up to 10~at.\% investigated here.
This implies that within our model the energy of the IS and FS coincide.
Distributing Re in the diffusion window yields a total of 7 configurations with $0-4$ Re atoms (6 of these are shown in Fig.~\ref{fig:diffusion_channels}), where combinations with two Re atoms result in three symmetry nonequivalent configurations.
%
\subsection{Diffusion barriers for Re, W, and Ta}
\label{subsec:solutebarriers}
\begin{figure}
        \includegraphics[width=0.35\textwidth]{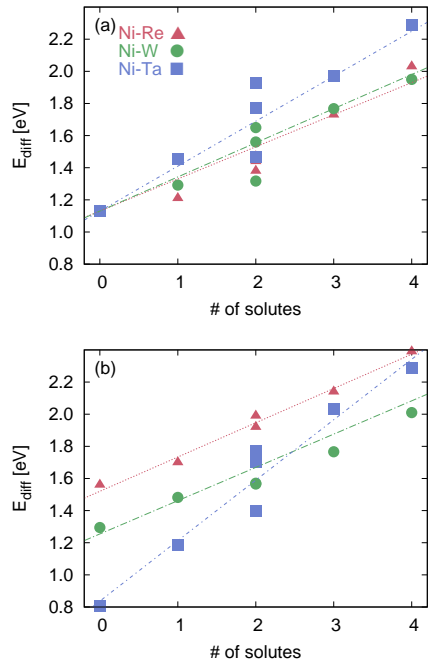}
	\caption{\label{fig:diffusionTaW} Diffusion barriers of (a) Ni and (b) solute atoms as a function of solute atoms in the diffusion window for Ni-Re (red triangles),  Ni-W (green circles), Ni-Ta (blue squares).
}
\end{figure}
Using DFT we have calculated the diffusion barriers of Ni and Re for all 7 possible configurations of the diffusion window model.  The diffusion barriers as a function of Re atoms in the diffusion window are shown in Fig.~\ref{fig:diffusionTaW} and the corresponding values are listed in Tab.~\ref{tab:barriers}.  We observe an almost linear increase in the diffusion barriers for both Ni (Fig. \ref{fig:diffusionTaW}(a), red triangles) and Re (Fig. \ref{fig:diffusionTaW}(b), red triangles).

In addition to Re we also performed DFT calculations for all configurations with $0-4$ solutes in the diffusion window in Ni-W and Ni-Ta.   
The results for the Ni and W diffusion barriers are given by green circles in Fig.~\ref{fig:diffusionTaW}.  The general trend is the same as in the Ni-Re system, the diffusion barriers increase with an increasing number of solute atoms in the diffusion window.  Interestingly, the effect of Re and W on the diffusion barriers of Ni is also quantitatively similar which is of importance for the mobility of Ni in these two alloys.  
The diffusion barriers in the Ni-Ta system are represented by blue squares in Fig.~\ref{fig:diffusionTaW}.  In pure Ni the diffusion barrier of Ta is lower than the one of Ni, but Ta in the diffusion window has a much larger effect on the diffusion barriers than Re or W.  This may be due to a simple size effect since Ta is the largest of the three solute elements and its presence in the diffusion window will narrow the path for the diffusing atom.  With only one Ta in the diffusion window the Ni diffusion barrier already increases to 1.45~eV.  Similarly, the diffusion barrier of Ta increases by almost 0.4~eV with a single Ta atom in the diffusion window. For the diffusion barriers with two solutes in the window we observe a large scatter depending on the occupation (see Fig. \ref{fig:diffusion_channels} (b) - (d)).

\begin{table}
	\caption{\label{tab:barriers} Diffusion barriers for Re, W, and Ta considering only the first NN of the transition state (in eV).  $0$ corresponds to no solute atoms in the diffusion channel. $1$, $3$, and $4$ correspond to configurations depicted in Fig. 3 (a), (e), and (f). The notation $2^{-}$, $2^{\setminus}$, and $2^{|}$ correspond to configurations shown in Fig. 2 (b), (c), and (d), respectively.
}
        \begin{ruledtabular}
        \begin{tabular}{l|ccccccc}
                window   & 0 & 1 & 2$^{-}$ & 2$^{\setminus}$ & 2$^{|}$ & 3 & 4 \\ \hline
                Ni in Ni-Re       &  1.13  &  1.21  & 1.38 & 1.47 & 1.44  &  1.73  & 2.03\\
                Re in Ni-Re       &  1.55  &  1.70  & 1.74 & 1.99 & 1.92  &  2.14  & 2.39\\[1ex]
                Ni in Ni-W        &  1.13  &  1.29  & 1.32 & 1.65 & 1.56  &  1.77  & 1.95\\
                W  in Ni-W        &  1.29  &  1.48  & 1.57 & 1.57 & 1.77  &  1.77  & 2.01\\[1ex]
                Ni in Ni-Ta       &  1.13  &  1.45  & 1.46 & 1.93 & 1.77  &  1.97  & 2.29\\
                Ta in Ni-Ta       &  0.81  &  1.19  & 1.40 & 1.70 & 1.80  &  2.03  & 2.29\\
        \end{tabular}
\end{ruledtabular}
\end{table}
The values listed in Tab.~\ref{tab:barriers} comprise the basic input data for the KMC simulations.

\section{Mobility in non-dilute alloys}
\label{sec:mobility}
KMC simulations were performed with $0-10$~at.\% solutes in 2~at.\% intervals.  In each KMC step the occupation of the diffusion window was evaluated and the corresponding diffusion barrier taken from Tab.~\ref{tab:barriers} was used to determine the rate constant via Eq.~\eqref{eq:htstrate}. 
For the diffusion processes with two solute atoms in the diffusion window  we take the average of all three possible diffusion barriers.
Simulations were run for temperatures of $500-1700$~K with $10^8 - 10^9$~KMC steps.
The vacancy diffusion coefficient is evaluated using Eq.~\eqref{eq:diffusion} to determine how the vacancy mobility is affected by the various solutes.  If the solute atoms were causing a significant slowdown of the vacancies this could retard vacancy flow needed in dislocation climb, which might contribute to the observed creep strengthening in Ni-based superalloys.

\subsection{Vacancy mobility}
\label{subsec:vacancy}
\begin{figure*}
        \includegraphics[width=\textwidth]{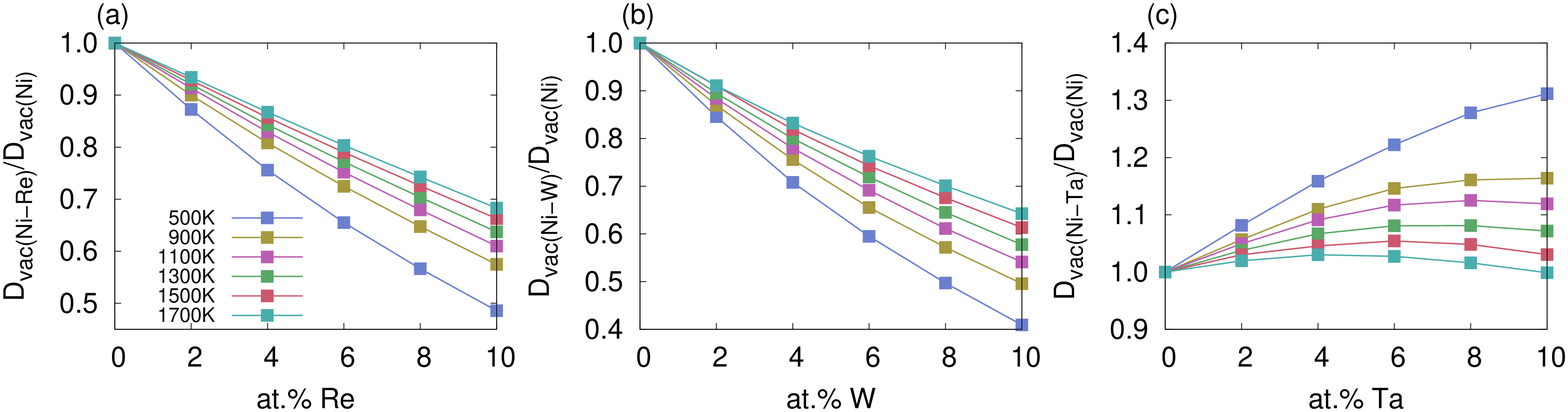}
        \includegraphics[width=\textwidth]{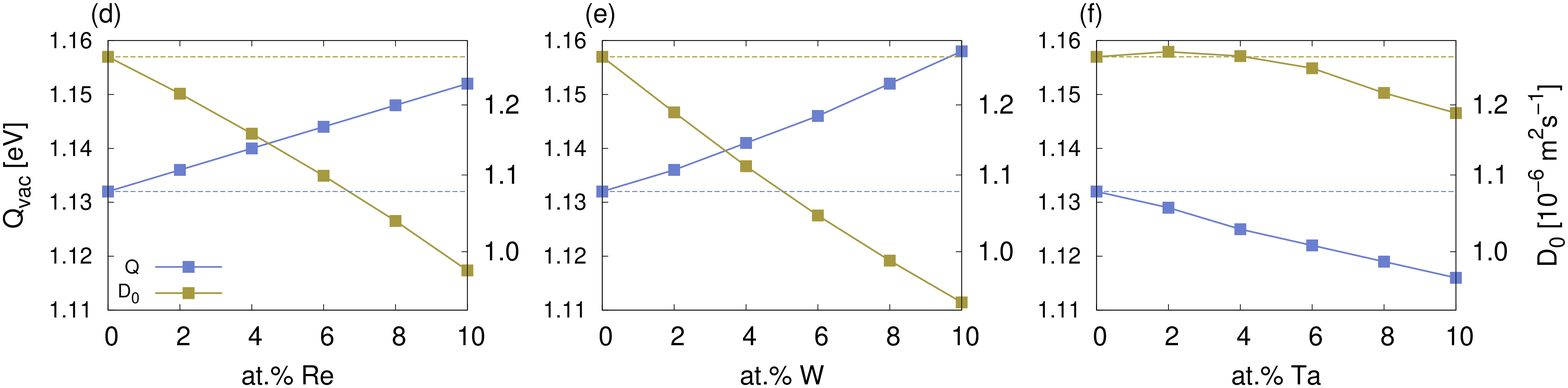}
        \caption{\label{fig:vacancy} Top:  relative slowdown of the vacancy mobility as a function of solute concentration for different temperatures for (a) Re, (b) W, and (c) Ta.  Bottom: effective activation energy $Q$ and the diffusion prefactor $D_0$ for vacancy diffusion for solute concentrations of $0-10$~at.\% for (d) Re, (e) W, and (f) Ta.
}
\end{figure*}
In Fig.~\ref{fig:vacancy}(a) the relative slowdown in vacancy mobility as a function of solute concentration is shown for different temperatures in the Ni-Re system.  The relative slowdown is given by the ratio of the vacancy diffusion coefficient in Ni + $x$~at.\% solute and in pure Ni.  Errors are in the range of $0.7-2.2$\% and the corresponding error bars are smaller than the symbol size.  In the case of Re the slowdown increases almost linearly with Re concentration, the corresponding slope increases with decreasing temperature.  
The change in vacancy mobility results from a combination of different effects that originate in the statistical interplay between host and solute diffusion.  Within our KMC model the rate constants for the atomistic diffusion processes vary depending on the chemical composition of the diffusion window, which makes an analytic description of the observed diffusivity even more complicated.
To aid the interpretation of the simulation results we have fitted the KMC data for the vacancy diffusion coefficients using Eq.~\eqref{eq:arrhenius} to extract the effective activation energy $Q$ and the prefactor $D_0$.  The change in $Q$ and $D_0$ with Re concentration is shown in Fig.~\ref{fig:vacancy}(d).  
In pure Ni without any Re atoms  the effective activation energy $Q=1.132$~eV corresponds to the self-diffusion barrier of Ni, and the diffusion prefactor $D_0 = 1.274\times10^{-6}$~m$^2$~s$^{-1}$ agrees well with the analytical value $D_0 = 1.230\times10^{-6}$~m$^2$~s$^{-1}$.
With increasing Re concentration the effective activation energy increases by $\sim 2$~meV/at.\% Re and the prefactor decreases by $\sim 2.5$\%/at.\% Re, both supporting the observed decrease in the vacancy diffusion coefficient.
Even though the change in the Ni and Re diffusion barriers with increasing number of Re atoms in the diffusion window is significant (Tab.~\ref{tab:barriers}), the change in the effective activation energy is rather small.  This can be understood when analyzing the distribution of processes obtained in the KMC simulation.  For all temperatures and Re concentrations the diffusion is dominated by Ni atoms without Re atom in the diffusion window.  For large Re concentrations and high temperatures Ni diffusion processes with one Re atom in the diffusion window contribute up to 20\%, whereas configurations with two or more Re atoms in the diffusion window play a minor role.
The relative decrease in the effective diffusion prefactor $D_0$ is much larger than the change in $Q$.  The decrease in $D_0$ is partially due to a decrease in the vacancy correlation factor $f_\text{vac}$ which decrease to about 0.87 for 10~at.\% Re and $T=1300$~K.
The vacancy slowdown in the Ni-Re system is  dominated at high temperatures by the decrease in the diffusion prefactor, with decreasing temperatures the effect is enhanced by the increase in the effective activation energy.  For Re concentrations of 10~at.\% and temperatures around $1300-1700$~K we observe a relative slowdown of $36-32$~\%, which is significant, but can only partially explain the Re-effect.  At the $\gamma$/$\gamma$'-interface and dislocations, it might be possible to achieve concentrations up to 20~at.\% Re due to segregation or dynamically through redistribution during dislocation climb. For such very high concentrations, we calculate a slowdown of $65-59$\% for a temperature range of $1300-1700$~K, which is closer to but still lower than the 75\% slowdown needed to account for the Re-effect.  In addition to a reduced vacancy mobility other effects are thus expected to contribute to the observed enhanced creep strength in Ni-based superalloys.

The results for Ni-W (Fig.~\ref{fig:vacancy}(b) and~(e)) are very similar to the Ni-Re system.  The diffusion barrier of W itself in Ni is lower than the Re diffusion barrier, but W reduces the vacancy mobility even more than Re.  This is due to the fact that within the investigated concentration range vacancy diffusion is dominated by the movement of Ni atoms, and W in the diffusion window increases the Ni diffusion barrier more strongly than Re.  This is also reflected in the slightly larger increase in the effective activation energy and larger decrease in the diffusion prefactor.  The relative slowdown at 10~at.\% W ranges from $42-36$~\% for temperatures of $1300-1700$~K.  Assuming again that a reduced vacancy mobility hinders dislocation climb and thus contributes to an increase in creep strength our results corroborate the fact that a similar effect as adding Re can be obtained by locally increasing the concentration of W in the $\gamma$-phase of Ni-based superalloys.~\cite{FMA15}  

The diffusion behavior in the Ni-Ta system is different from Ni-Re and Ni-W.  The main reason for this is the low diffusion barrier of Ta in Ni.  By adding a fast diffuser to the Ni matrix we would expect an increase in the vacancy mobility.  As shown in Fig.~\ref{fig:vacancy}(c) this is only partially true and more pronounced for low temperatures.  The negative slowdown or rather speed-up  of the vacancies is reduced by the strong increase in the Ni and Ta diffusion barriers with additional Ta in the diffusion window (Tab.~\ref{tab:barriers}).  This is also reflected by the change in $Q$ and $D_0$ as a function of Ta concentration, Fig.~\ref{fig:vacancy}(f).
The diffusion prefactor hardly changes up to 6~at.\% Ta, followed by a slight decrease of 8~\% for 10~at.\% Ta, and the effective activation energy decreases by $\sim 1.6$~meV/at.\% Ta.
At high temperatures the change in vacancy mobility as a function of Ta concentration is governed by the diffusion prefactor, whereas  at lower temperature the speed-up is enhanced by the decrease in the effective activation energy.  The overall shape of the curves reflects the change in trend in $D_0$ as well as the competition between a decrease in $D_0$ and a decrease in $Q$.  At temperatures of $1300-1700$~K the overall speed-up is less than 8~\%, i.e. the effect of Ta on the vacancy mobility is small.

\subsection{Ideal solid solutions}
\label{subsec:idealsolid}
For a binary random alloy without any interactions the kinetic transport coefficients $L_{ij}$ of the Onsager matrix can be determined analytically~\cite{MAA89} and it has been shown that these agree well with numerical results of KMC simulations.~\cite{VYC10}
Together with the thermodynamic factor matrix $\tilde{\Theta}$ the matrix of diffusion coefficients $\mathbf{D}$ is given by~\cite{VYC10}
\begin{equation}
\label{eq:diffmatrix}
\begin{pmatrix}
 D_{AA}  & D_{AB} \\
 D_{BA}  & D_{BB}
\end{pmatrix}
=
\begin{pmatrix}
 \tilde{L}_{AA}  & \tilde{L}_{AB}  \\
 \tilde{L}_{BA}  & \tilde{L}_{BB}
\end{pmatrix}
\begin{pmatrix}
 \tilde{\Theta}_{AA}  & \tilde{\Theta}_{AB}  \\
 \tilde{\Theta}_{BA}  & \tilde{\Theta}_{BB} 
\end{pmatrix}
\quad ,
\end{equation}
where $\tilde{L}_{ij} = L_{ij} \Omega k_\text{B}T$ and $\Omega$ is the volume of a substitutional site.
Again for an ideal solid solution the thermodynamic factor matrix can be calculated analytically from the concentrations of the alloy components and the vacancies.~\cite{VYC10}
The atomistic and analytical expressions for the thermodynamic factor matrix and the kinetic transport coefficient are reviewed in Appendix~\ref{appendix}.
The eigenvalues of the diffusion coefficient matrix can be interpreted in terms of relevant diffusion processes.  In particular it has been shown that for ideal solid solutions and in the limit of small vacancy concentrations the larger eigenvalue can be associated with the vacancy diffusion coefficient and the smaller one with the intermixing coefficient.~\cite{KBR89}  

\begin{figure*}
        \includegraphics[width=\textwidth]{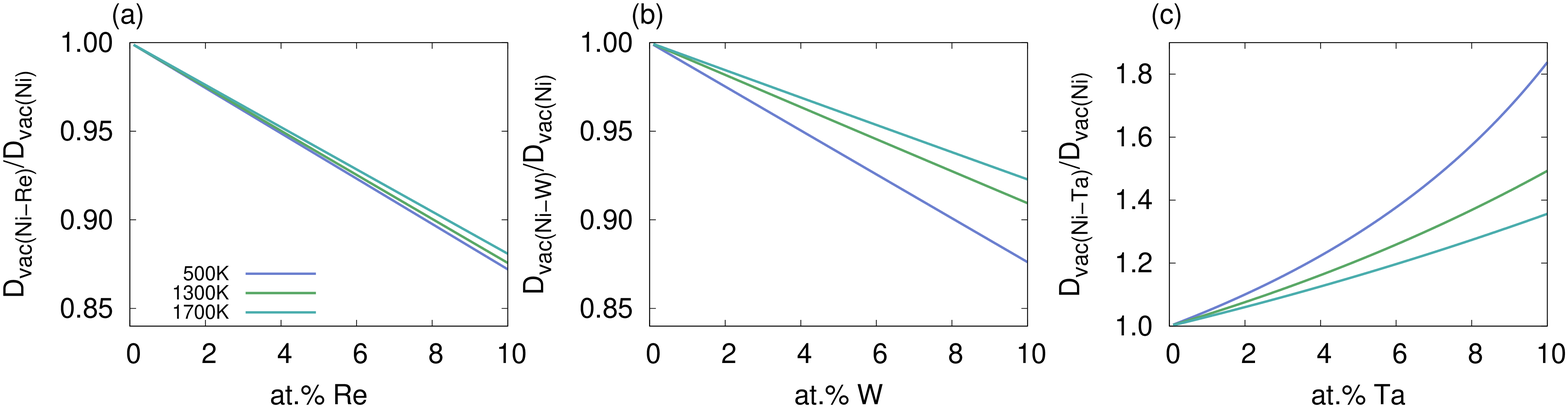}
        \caption{\label{fig:analytical} The predicted, relative slowdown of the vacancy mobility as a function of solute concentration for different temperatures for (a) Re, (b) W, and (c) Ta for an ideal random alloy. The slowdown is underestimated as compared to our KMC simulations.} 
\end{figure*}

We have evaluated the diffusion coefficient matrix in Eq.~\eqref{eq:diffmatrix} for the Ni-Re, Ni-W, and Ni-Ta system using the diffusion barriers in pure Ni given in the Tab.~\ref{tab:barriers}.  From the eigenvalues of $\mathbf{D}$ we have determined the corresponding vacancy diffusion coefficients as a function of solute concentration.  The corresponding slowdown in vacancy mobility is shown in Fig.~\ref{fig:analytical}.  Qualitatively we observe the same behavior as a function of concentration and temperature as for our KMC model, but quantitatively there are significant differences.  
For 10~at.\% Re and temperatures of $T=1300-1700$~K the slowdown is only around 12\% in the ideal solid solution (Fig.~\ref{fig:analytical}(a)) compared to $36-32$\% when considering the influence of Re in the diffusion window on the diffusion barriers (Fig.~\ref{fig:vacancy}(a)).  This increase by a factor of three can be attributed to a change in the distribution of diffusion barriers encountered during the KMC simulations.

The differences are even more pronounced for Ni-W.  In a random alloy the vacancy slowdown for 10~at.\% W and $T=1300$~K is only about 9\%, i.e. the effect is smaller than for Re.  This is expected since in the ideal solid solution only the diffusion barriers in pure Ni enter the analytical model and here the diffusion barrier of W is smaller than the one of Re.  In our KMC model the vacancy slowdown is larger in the Ni-W system.  As discussed in Sec.~\ref{subsec:vacancy} this is due to the larger influence of W on the Ni diffusion barriers, which cannot be captured by the analytical model.

For Ni-Ta the analytical model predicts a monotonic increase of the vacancy mobility up to 10~at.\% Ta which is reasonable when adding a fast diffuser.  The effect is much larger than in our KMC model and the change in trend, i.e. the observed decrease in vacancy mobility for larger Ta concentrations, is missing.  Again we can clearly identify the influence of additional Ta on the diffusion barriers as the  source for the discrepancy between the KMC simulations and the analytical model.

From the comparison of the KMC results with the analytical model for a binary random alloy it clearly follows that the change in diffusion barriers due to the presence of solute atoms strongly influences the composition and temperature dependence of the vacancy mobility.

\section{Conclusions}
\label{sec:conclusion}

Using a combined DFT and KMC study we have investigated the effect of solute concentration on the vacancy mobility in binary Ni-Re, Ni-W, and Ni-Ta alloys.  Adding a slow diffuser (Re, W) decreases the vacancy mobility, whereas a fast diffuser (Ta) increases the vacancy mobility.
Within our KMC simulations we take into account the effect of solute atoms on the microscopic diffusion barriers through the so-called window model, i.e. we determine the dependence of the diffusion barriers on the chemical composition of the diffusion window.  Within this simple interaction model we observe significant deviations from the vacancy diffusion within an ideal solid solution.  For the slow diffusers Re and W the KMC simulation predict a much larger vacancy slowdown, whereas for the fast diffuser Ta the speed-up is reduced compared to a random alloy.

These rather notable quantitative differences originate solely from the influence of the solute atoms on the diffusion barriers.  In particular the fact that within the KMC simulations the vacancy slowdown is stronger in Ni-W than in Ni-Re emphasizes the importance of including these interactions: the magnitude of the slowdown is primarily determined by the effect of the solute atom on the host diffusion barriers, rather than the diffusion barrier of the solute itself 

In the context of Ni-based superalloys it becomes evident that local Re concentrations of up to 10~at.\% considerably slow down vacancy diffusion, but not sufficiently to be the sole cause for the observed Re-effect.  To obtain the expected slowdown, the local Re concentration needs to exceed $20$~at.\% Re which might be possible at dislocations and the $\gamma$/$\gamma$'-interface.  
The KMC simulations also confirm that a similar effect could be achieved by adding W, where again the slowdown is determined by the effect of the solute on the host diffusion barrier instead on the diffusion barrier of the solute itself. 

Based on our results of the KMC simulations it becomes apparent that within non-dilute binary alloys a detailed treatment of solute-solute, solute-host, and solute-vacancy interaction energies is crucial for the correct prediction of atomic diffusion.


\begin{acknowledgments}
The authors acknowledge funding by the Deutsche Forschungsgemeinschaft (DFG) through project C2 of the collaborative research centre SFB/TR 103 "From Atoms to Turbine Blades - a Scientific Approach for Developing the Next Generation of Single Crystal Superalloys."
\end{acknowledgments}


\appendix
\section{Calculation of Diffusion Matrix}
\label{appendix}
We briefly review  the expressions for the calculation of the diffusion matrix Eq.~\eqref{eq:diffmatrix} using the thermodynamic factor matrix and the transport coefficients, which can be determined numerically and analytically for the case of an ideal solid solution.
A detailed derivation and discussion of these expressions is given in Ref.~\onlinecite{VYC10}.\\
The thermodynamic factor matrix  $\tilde{\Theta}$ can be determined numerically using grand canonical Monte Carlo simulations with a fixed number of crystal sites $M$ by
\begin{equation}
        \tilde{\Theta} = \frac{M}{Q}\left(
        \begin{matrix}
                \left<\Delta^{2}_{jj}\right> & -\left<\Delta^2_{ij}\right>\\
                -\left<\Delta^{2}_{ij}\right> & \left<\Delta^2_{ii}\right>
        \end{matrix}
\right)\quad,
\end{equation}
where 
\begin{equation}
        \left<\Delta^2_{ij}\right> = \left<N_iN_j\right>-\left<N_i\right>\left<N_j\right>,\quad \text{with  }i,j = A,B\quad,
\end{equation}
and
\begin{equation}
        \begin{split}
                Q = & \left(\left<N^2_i\right>-\left<N_i\right>^2\right)\left(\left<N^2_j\right>-\left<N_j\right>^2\right)\\
                    & -\left(\left<N_iN_j\right>-\left<N_i\right>\left<N_j\right>\right)^2\quad.
        \end{split}
\end{equation}
The angular brackets denote ensemble averages and $N_{i,j}$ is the number of atoms $A$ and $B$. 
For an ideal solution the thermodynamic factor matrix can be expressed analytically as
\begin{equation}
        \tilde{\Theta}=\left(
        \begin{matrix}
        \frac{(1-x_j)}{x_ix_V} & \frac{1}{x_V}\\
        \frac{1}{x_V} & \frac{(1-x_i)}{x_jx_V}
        \end{matrix}
\right)\quad,
\end{equation}
where $x_{i,j}$ is the concentration of element $i,j =A,B$ and $x_V$ for vacancies $V$.\\
The kinetic transport coefficients $L_{ij}$ can be calculated numerically from the displacement of the atoms, using e.g.  KMC simulations, with
\begin{equation}
        L_{ij} = \frac{1}{\Omega k_{\rm B}T}\tilde{L}_{ij}\quad,
\end{equation}
where $\Omega$ is the volume per substitutional site and
\begin{equation}
        \tilde{L}_{ij} = \frac{\left<\left(\sum_{\xi}\Delta\vec{R}^i_{\xi}(t)\right)\left(\sum_{\xi}\Delta\vec{R}^j_{\xi}(t)\right)\right>}{(2d)tM}\quad.
\end{equation}
$\Delta\vec{R}^i_{\xi}$ denotes the vector linking the end points of the trajectory of atom $\xi$ of atomic species $i$ after time $t$, $d$ is the dimensionality of the system and $M$ denotes the number of sites in the crystal.\\
In an ideal random alloy, the analytical expression for $\tilde{L}_{ij}$ with different rate constants for two elements $k_{A}\neq k_{B}$ is
\begin{equation}
        \begin{split}
                \tilde{L}_{ii} & = x_Vx_i\rho a^2k_i\left(1-\frac{2x_jk_i}{\Lambda}\right)\quad,\\
                \tilde{L}_{ij} & = \frac{2\rho a^2k_{i}k_{j}x_Vx_ix_j}{\Lambda}\quad,
\end{split}
\end{equation}
where
\begin{widetext}
\begin{equation}
        \Lambda = \frac{1}{2}\left(F+2\right)\left(x_ik_i+x_jk_j\right) -k_i-k_j+2\left(x_ik_j+x_jk_i\right) +\sqrt{\left(\frac{1}{2}\left(F+2\right)\left(x_ik_i+x_jk_j\right)-k_i-k_j\right)^2+2Fk_ik_j}\quad,
\end{equation}
\end{widetext}
and
\begin{equation}
        F=\frac{2f}{1-f}\quad .
\end{equation}
 $f$ is the correlation factor for a single element with the crystal structure of the $A-B$ alloy.\\
Using these expressions for the thermodynamic factor matrix and the kinetic transport coefficients, we can calculate the diffusion matrix in Eq.~\eqref{eq:diffmatrix}.

\end{document}